\documentclass[aps,prc,twocolumn,balancelastpage]{revtex4-1}
\usepackage[utf8x]{inputenc}
\usepackage{amsmath,amsfonts,verbatim,graphicx,epsfig,rotating,appendix,srcltx}
\usepackage{float}

\def\fm3{\rm ~fm^{-3}}
\def\MeV{\rm ~MeV}	
\newcommand{\beq}{\begin{equation}}
\newcommand{\eeq}{\end{equation}}
\newcommand{\beqa}{\begin{eqnarray}}
\newcommand{\eeqa}{\end{eqnarray}}

\graphicspath{{eps/}}

\begin{document}

\title{Systematic analysis of symmetry energy effects  in the neutron star crust properties}
\author{S.Kubis}
\author{D.E. Alvarez-Castillo}
\affiliation{H.Niewodnicza\'nski Institute of Nuclear Physics, Radzikowskiego 152, 
31-342 Krak\'ow, Poland}

\begin{abstract}

The functional form of the nuclear symmetry energy in the whole range of
densities relevant for the neutron stars is still unknown. Discrepancies concern
both the low as well as the high density behaviour of this function. By use of
B\'ezier curves three different  families of the symmetry energy shapes, relevant
for different density range were introduced.  Their  consequences for the
crustal properties of neutron stars are presented.

\end{abstract}

\maketitle

\section{Introduction}
The basic quantity in the description of infinite nuclear matter filling out the
interior of neutron star is the energy per particle expressed in term of baryon
number  density $n=n_p+n_n$ and isospin asymmetry $\alpha = \frac{n_n-n_p}{n}$
of the system:
\beq
E(n,\alpha)=V(n) + E_s(n)\, \alpha^2 + {\cal O}(\alpha^4)
\label{Enuc}
\eeq
Instead of $\alpha$ it is useful to use proton fraction $x$, and then $\alpha =
(1-2x)$. Here we assumed the only constituents of stellar matter are nucleons and
leptons: electrons and muons. Around and above the nuclear density $n_0 = 0.16 \fm3$
nucleons and leptons form a quantum liquid, which stands for liquid core of the
neutron star. Slightly below $n_0$ matter cannot exist as a homogeneous fluid -
the one-phase system is unstable and the coexistence of two phases is required.
At these densities matter clusterizes  into positive nuclei immersed in a
quasi-free gas of neutrons and electrons. Model calculations show this type of
matter forms a Coulomb lattice with solid state properties and corresponds to the
crust covering liquid core of a star. The presence of the crust is affirmed by
the glitching phenomenon  observed for some pulsars.  For typical NS masses,
between 1-2 $\textit{M}_\odot$, the most of the stellar matter is occupied by the core,
so the global parameters like the mass, radius, moment of inertia are completely
determined by the functional form of  the Eq.~(\ref{Enuc}). Whereas the isoscalar
part $V(n)$ corresponds mainly for the stiffness of Equation of State (EOS)
which is relevant for the maximum mass of NS,  the isovector part $E_s(n)$ is
responsible for the chemical composition of the matter. Through the
$\beta$-equilibrium equations
\beq 
4(1-2x)E_s(n) = \mu_n - \mu_p = \mu_e = \mu_\mu
\label{beta}
\eeq
the proportions of all particles are determined. The symmetry energy is also
relevant for the crust-core transition in NS as it was shown
in~\cite{Kubis:2006kb}. 
It's role can be explicitly seen if one looks at the compressibility under
constant chemical potential relevant for the stability of homogeneous beta
equilibrated  nuclear matter:
\beq
 K_\mu = n^2(E_s''\alpha^{2} +V'') + 2 n (E_s'\alpha^{2} +V') -
\frac{2 \alpha^{2} E_s'^{2} n^{2} }{E_s},
\label{Kmu}
\eeq
and when $K_\mu >0$ the matter is stable.
This formula accounts for the bulk approximation and can be improved by
inclusion of finite size effects like Coulomb and surface contributions. Such
corrections were studied in~\cite{1971NuPhA.175..225B} were it was shown that  
\begin{equation}
v(Q)=v_{min}=v_0+2(4\pi e ^{2} \beta)^{1/2}-\beta k_{TF}^{2}
\end{equation}
is the minimal value for stable density modulations for the $Q$ momentum.
Stability of matter is given by the condition $v(Q)>0$. Note that the vanishing
of $v_0$ is equivalent to the vanishing of $K_{\mu}$ since they are related by
\begin{equation}
 v_0(n)=\frac{8K_{\mu}(n)E_s(n)}{n^{2}}\left(\frac{\partial \mu_n} {\partial n_n} \right)^{-1}.
\end{equation}
Both of these approaches consider stability of a one phase system against
density fluctuations. However there exists another approach based on treating
the NS crust  as a two component system subject to the Gibbs conditions for
mechanical and chemical stability expressed by:
\begin{equation}
 p^{I}   =  p^{II} ~~~~~,~~~~~
\mu_n^{I}   =  \mu_n^{II}~~~~~,~~~~~\mu_e^{I}   =  \mu_e^{II}.
\label{coex}
\end{equation}
Where the first component (I) corresponds to clusters composed of protons and
neutrons immersed in a pure neutron liquid (II) component. Both phases are
permeated by  degenerated electrons. As density increases towards the star
interior the two phase system can no longer exist and it signals the crust-core
transition. These approaches correspond to three different critical densities
$n_{c}(K_{\mu})$, $n_{c}(Q)$, and $n_{c} (1\leftrightarrow 2)$ which will be
presented for various models in the following sections.

In this work we are interested in the crustal properties, so it seems natural to
focus on the symmetry energy form around saturation density $n_0$, to which the 
critical density $n_c$ is closely located. However, as was suggested in
\cite{kubisAPPB41} the crust is affected by the star compactness which depends
on the $E_s$ form at densities much higher than $n_0$. Furthermore recent
experimental measurements \cite{2010PhRvL.104t2501N}  show non-standard behaviour of $E_s$ at
densities much below $n_0$.  All these issues lead us to the idea  to somehow
"factorize" the shape of $E_s$ to see how symmetry energy
at different  ranges of density affects the NS crust.
It appeared to be possible to define models in which the $E_s$ is changing at
chosen range of density whereas the rest of its shape is kept the same.
We have constructed the three main families of models with varying shape
at very low, $n \rightarrow 0$, intermediate $n\approx n_0$ and very high densities
$n \gg n_0$. The work is organized as follows: in the section II we present
recent experimental data useful in constraining the form of the symmetry energy,
in the section III the construction of different models is shown. In the last
two sections the results of the calculation and the astrophysical constraints
coming form NS observations are discussed.

\section{Measured symmetry energy properties in laboratory experiments}

The symmetry energy $E_s$ properties around saturation density can be 
inferred from laboratory experiments. For nuclei, it corresponds to the
volumetric symmetry energy  term $S_v$ in the liquid droplet
model~\cite{1974AnPhy..84..186M}: $E_s(n_0)=S_v$, which also contains the
surface contribution to the symmetry energy $S_s$. By measuring in nuclei these
two quantities is possible to derive $E_s(n_0)$ and $L$  which is related to its
slope 
\begin{equation}
L=3n_{0}\frac{\partial E_s(n)}{\partial n}\bigg|_{n_0}.
\end{equation}
Not precisely available at the moment from experiments but also relevant for NS matter is the curvature of the symmetry energy:
\begin{equation}
 K_{s} = 9n_{0}^2 \left. \frac{\partial^2E_{s}(n)}{\partial n^2}\right|_{n_0}.
\label{Compressibility}
\end{equation}
Neutron rich nuclei feature halos composed mainly of neutrons while most of the
protons stay in nuclear cores. Such neutron skin thickness is dependent on both
volumetric and surface symmetry energies (by means of the coefficient $S_v/S_s$)
and provides another way of measuring $E_s(n_0)$. $S_v$ itself can be extracted
from heavy ion collisions where isospin diffusion occurs: two colliding nuclei
with opposite neutron abundances exchange components reaching isospin
equilibration. Both, neutron skin thickness measurements \cite{Chen:2010qx} and isospin
diffusion \cite{Tsang:2008fd} shows that $L$ is placed in the range between 40
and 80 MeV. 

Experiments allows for measurements of asymmetric matter compressibility 
$K_{asy}$ being related to $K_s$ by the approximate relation $K_s \approx
K_{asy}+6 L$. Recent experimental analysis 
\cite{Chen:2004si,Centelles:2008vu,Li:2007bp} shows that $K_{asy}$ 
takes values between -650 and -400 MeV. It means that, including the
discrepancies of $L$, one may estimate that $K_s$ is in a broad range
between -400 and 100 MeV. 

The density around $n_0$ is the natural place where
the efforts in determining of symmetry energy behavior are focused.
However even if the values of slope $L$ or the curvature $K_s$  achieve 
satisfactory accuracy the symmetry energy shape remains uncertain in both very
low and very high density which are relevant for NS properties. 
For some years, the very  low density  part of the nuclear energy has
been studied as well \cite{Kowalski:2006ju,2010PhRvL.104t2501N}.
In these  works the  symmetry energy is probed at $n \approx 10^{-2} \fm3$
and appears to take large values  $\sim 10$~MeV at this range of density.
The result is very interesting. 
First, it allows for going  beyond the poor and not very
restrictive expansion around the $n_0$, and secondly it is highly relevant for
the crust-core transition in neutron star.  

\section{B\'ezier Curves for the Symmetry Energy}
For a good description of the neutron star properties it is important to know
the symmetry energy functional form in the whole range of densities. From the
aforementioned measurements one can pin down the symmetry energy properties
around saturation but both low and high density parts are not well determined.
Therefore there is large freedom to assume the behavior of $E_s(n)$ and
introduce different parametrization.

Previously used parametrization (often polynomials) were not convenient since
once the same saturation point properties are fulfilled the values  far from
saturation are fixed, thus sometimes leading to unphysical results, like
unstable EOS (MDI parametrization) for large negative vales  of $E_s$ at high
densities~\cite{kubisAPPB41}. Polynomial interpolation presents some advantages
as it may produce any shape but  might present problems when derivatives are
computed. At the boundary of the domains of interpolation the analicity is
lost.  In particular for pressure and compressibility the first and second
derivatives are required and such polynomial interpolation leads to artifacts in
the EOS.
\begin{figure}[ht!]
\center
{\includegraphics[width=.7\columnwidth]{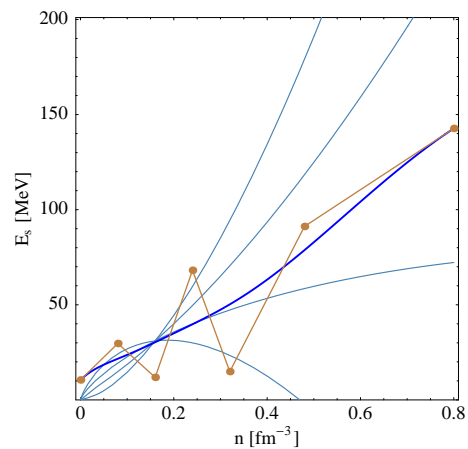}}
\caption{B\'ezier Curve (thick blue) for seven control points (brown). The thin lines represent symmetry energies for the MDI models shown as a reference.}
\label{bezier}
\end{figure}
On the contrary B\'ezier curves~\cite{BezierCurves} allow for construction of
different $E_s$ shapes that respect saturation properties,  like $E_s(n_0)$ and
$L$, but can take any value at low and high density values, 
see Fig.~\ref{bezier}. The
resulting B\'ezier curve is then an analytical function that offers big
advantages over interpolated functions.

B\'ezier curve of degree n is based on $n+1$ control points $\textbf{P}_i$,
where $i=0,1,2\ldots n$.  For implementation purposes, an arbitrary number of
control points can be chosen to cast the curve shape as desired by means of the
following relation:
\begin{equation}
 \textbf{B}(t)=\sum^{n}_{i=0}{n \choose i}(1-t)^{n-i}t^{i}\textbf{P}_i, \qquad  t \in [0,1]
\end{equation}
where ${n \choose i}$ is the binomial coefficient. 

The constructed models for $E_s$ are presented in the following subsections and
their control points are tabelarized in the appendix.
As we are interested only in the symmetry energy effects,
the isoscalar part $V(n)$ in the Eq.(\ref{Enuc}), required for the  of the full
 nuclear model, is kept all the time the same. We used the functional form taken 
from \cite{1988PhRvL..61.2518P} for which the symmetric matter 
compressibility is $K_0 = 240$~MeV.

\subsection{Low density symmetry energy effects and the neutron clusterization problem}

\begin{figure}[htbp!]
\centering
\includegraphics[width=.8\columnwidth]{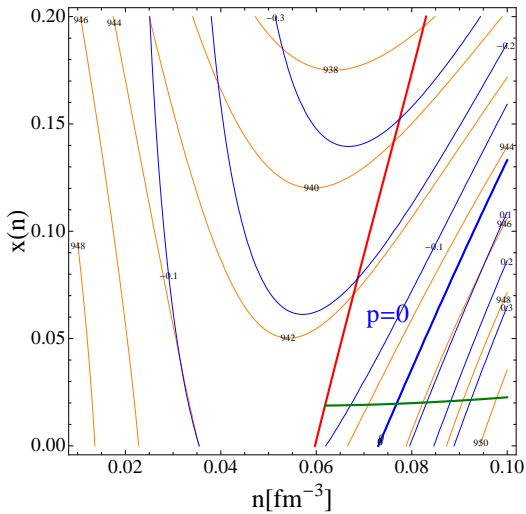} 
\caption{$\Delta E \!- \!k10$ model in which $E_s(0)=10\MeV$. Clusterization of
pure neutron matter occurs when the energy per baryon is given by $\tilde{E}(n,x)$.}
\label{kmodelsDiagram}
\end{figure}

In an earlier work the authors implemented a set of models motivated by the low
density behaviour of the symmetry energy that were introduced to explore NS
properties~\cite{kubisAPPB41}. However a  detailed analysis of phase transitions
revealed pathological properties.  Fig.~\ref{kmodelsDiagram} shows the phase
diagram of a model similar to the earlier $k10$ in which $E_s(n\sim 0)=10$~MeV
and for which the thermodynamical Gibbs  conditions hold.   In it, the region to
the left of the  spinodal line (in red) where the proton fraction is zero (pure
neutron matter) is reached by isobaric lines of negative pressure which
indicates clusterization of neutrons. 
This is in disagreement with common
knowledge that neutrons never clusterize. Here we propose a method avoiding this
pathology. In order to keep  finite values of $E_s$ at vanishing density 
it is then necessary to correct the whole expression for the 
total energy of nuclear matter.
First, we introduce the following symmetry energy expression:
\begin{displaymath}
E_s(n) = \left\{ \begin{array}{ll}
                     E_s^{PALu}(n) +E_s^{B\acute{e}zier,k}(n) & \textrm{if $n<n_0$}\\
		     E_s^{PALu}(n) & \textrm{if $n>n_0$},
                 \end{array} \right.
\end{displaymath}
where $E_s^{PALu}(n)$ is the symmetry energy introduced in the work 
\cite{1988PhRvL..61.2518P} with the interaction part $F(u)=u$.
$E_s^{B\acute{e}zier,k}(n)$ is a B\'ezier curve that modifies the low density
part in such a way that $E_s^{B\acute{e}zier,k}(0)+E_s^{PALu}(0)=k$, $k$ being
any value in MeV. In this way, $E_s(n)$ can take the $k=E_s(0)$ value for which
the corresponding B\'ezier curve is designed for. With it,  the energy per
particle becomes:
\begin{equation}
 \tilde{E}(n,x)=V(n)+(1-2x)^{2}E_{s}(n).
\label{EnPAL}
\end{equation}
The above formula still needs to be corrected for neutron clusterization. 
In order to \textit{avoid neutron clusterization}, the following correction is defined:
\begin{displaymath}
 \Delta E(n) = \left\{ \begin{array}{ll}
                    E_s^{B\acute{e}zier}(n) & \textrm{if $n<n_0$}\\
		     0 & \textrm{if $n>n_0$,}
                 \end{array} \right. 
\end{displaymath}
so that the final form of the energy is given by
\begin{equation}
 E(n,x)=\tilde{E}(n,x)-\Delta E(n).
\end{equation}
\begin{figure}[htbp!]
\center
{\includegraphics[width=0.4\textwidth]{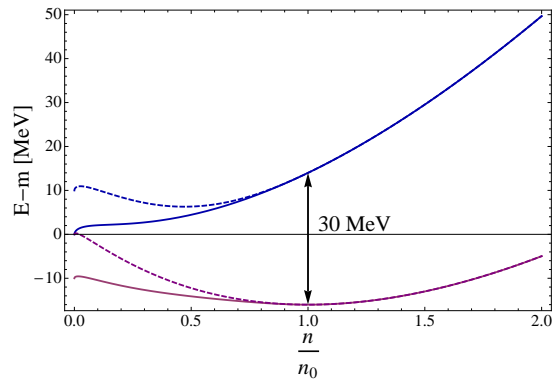}}
\caption{Energy per particle $E(n,x)$ for the $\Delta E${\it -k10} model with correction (solid lines) and without it: $\tilde{E}(n,x)$(dashed lines). The blue lines
 correspond to pure neutron matter while the purple ones correspond to symmetric nuclear matter.}
\label{EsymDefPALuDeltaE}
\end{figure}
We call $\Delta E${\it -k} models when the above modifications are applied. In fact, the symmetry energy can be recovered by definition:
\begin{equation}
 E_{s}(n)=E(n,x=0)-E(n,x=\frac{1}{2}).
\label{EnplusDeltaEsym}
\end{equation}
and captures the features of the $k$ models (finite symmetry energy values at almost zero baryon density). At saturation point these models have the same 
values
as the PALu model (in MeV):
\begin{equation}
 E_s(n_0)=30, \qquad L=77, \qquad  K_s=-26 \nonumber.
\end{equation}
Let us recall that the $E_s^{B\acute{e}zier}$ curves where defined in such a way
to preserve the known properties of symmetric nuclear matter, i.e., its
compressibility at saturation and its minimum. These conditions are fulfilled
because the following derivatives with respect to baryon number density vanish
at saturation:

\begin{eqnarray}
 \Delta E'(n_0)=0 ,\nonumber\\
 \Delta E''(n_0)=0.
\end{eqnarray}
\begin{figure}
\centering
 \includegraphics[width=.8\columnwidth]{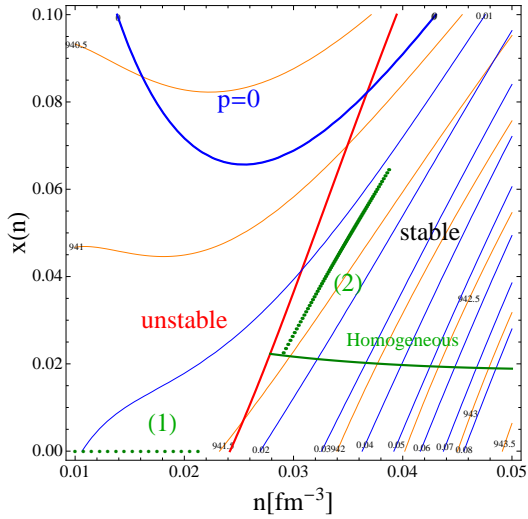}

\caption{$\Delta E${\it -k10} model in which $E_s(0)=10\MeV$. Clusterization of
pure neutron matter is now impossible for this model when the $\Delta E$ 
correction is included. Close to the spinodal (in red), the system splits into
two phases, (1) - pure neutron (2) - nuclear matter represented by dotted green
lines.}

\label{DeltaEkmodelsDiagram}
\end{figure}
\begin{figure}[hb!]
\centering
\includegraphics[width=7cm]{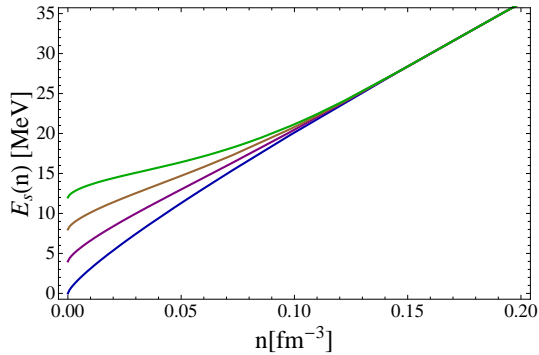}
\caption{Nuclear symmetry energy for $\Delta E${\it -k} models. The curves 
correspond to $E_{s}(0) = 0,4,8,12$~MeV.}
\label{Es-DeltaEk}
\end{figure}

\subsection{Different symmetry energy slopes at saturation density}

\begin{figure}[!htpb]
\includegraphics[width=7cm]{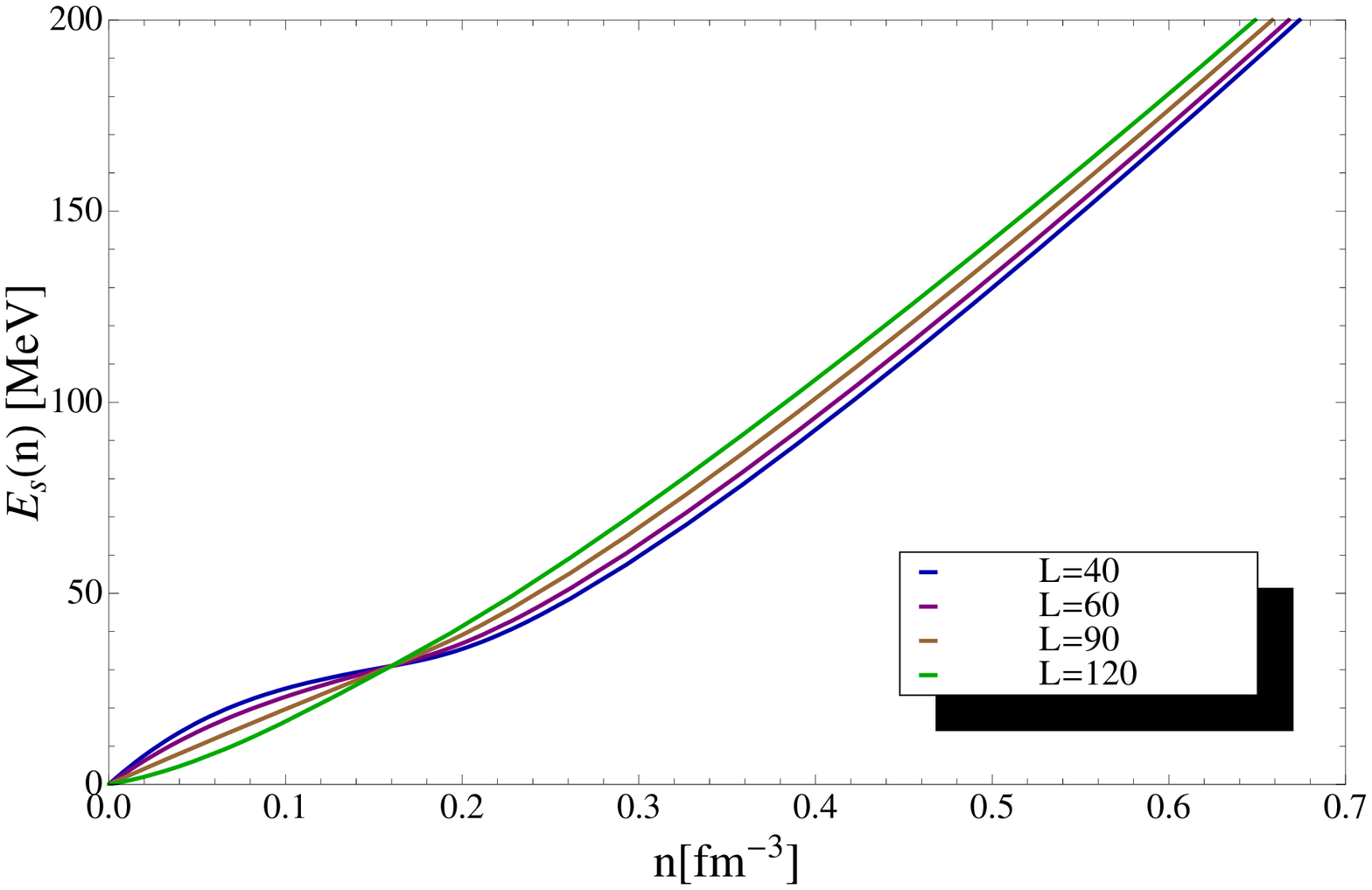}
\includegraphics[width=7cm]{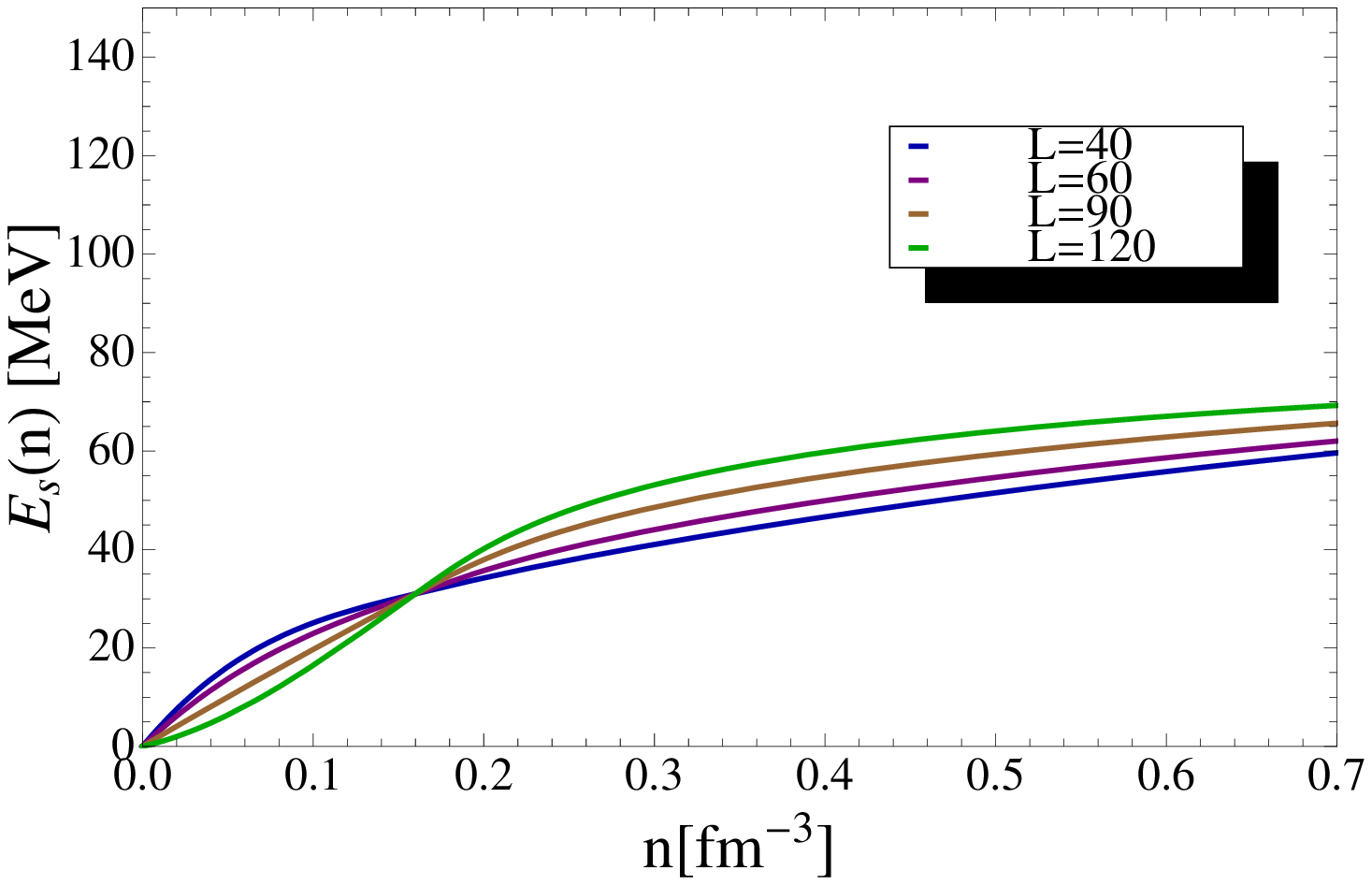}
\caption{Different symmetry energy shapes for the {\it L-high-c} models (top)
and for  {\it L-high-b} (bottom).}
\label{EsLvar-DU}
\end{figure}

The importance of the slope of the symmetry energy for the NS crust-core
transition point can be easily seen in Eq.~(\ref{Kmu}). The last term in that
equation  is the one responsible for the stability breaking and includes the
square of the symmetry energy slope.  The measured value of $L$ will influence
the behaviour of $E_s$ away from saturation. In this section we introduce two
families of models with the same symmetry energy forms at high and low densities
but with variable slope at saturation point. There at saturation, they share the
common values
\begin{equation}
E_s(n_0)=31, \qquad  K_s=0 \nonumber.
\end{equation}
whereas the slope $L$ takes values in the range between 40-120~MeV corresponding
to those reported by various experiments \cite{Chen:2007ih}.
These properties are established by joining two B\'ezier curves describing the
low and high density parts joint at saturation point with continuous derivatives
up to second order. At low densities all of them go to zero while at high
densities they follow the behavior of either b or c  MDI models~\cite{kubisAPPB41}.
The two families are presented in Fig.~\ref{EsLvar-DU}. The reason of
introducing the b or c-like behaviour at high densities comes from the analysis
of direct URCA constraint which is discussed in detail in the subsection D.
\subsection{High density region symmetry energy effects}
\begin{figure}[ht!]
\center
{\includegraphics[width=7cm]{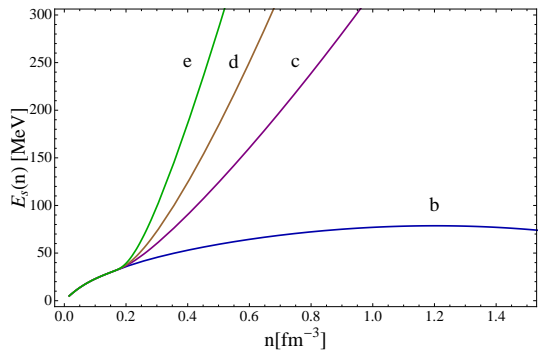}}
\caption{Symmetry energies for the {\it high-$E_{s}$} models.}
\label{EsHighValues}
\end{figure}

The critical density for the crust-core transition is the most relevant quantity
for determination of the NS crust properties. Its value determines the crust
thickness and moment of inertia. The crust-core position depends only on the
behavior of the symmetry energy below saturation. However the crust thickness is
not only determined by the critical density but  also by the global parameters
of the NS like its mass and radius. Those global parameters
 depend on the high density part of $E_s$. To explore such effects we construct a set of
models with the same low density part implying the same crust-core transition
but with different high density part. In relation to that, the authors of
~\cite{2001ApJ...550..426L} derive an approximate formula assuming a polytropic
equation of state for a thin, light crust but deviations from those assumptions
can lead to completely different results and are due to the high density 
$E_s$ region, as will be presented in the following sections. 
We introduce the {\it high-$E_{s}$} models which have the following values (in
MeV):
\begin{equation}
 E_s(n_0)=31, \qquad  L=60, \qquad  K_s=0 \nonumber
\end{equation}
and are composed of two B\'ezier curves describing both the low and high density
parts joint at saturation point with continuous derivatives, just like in the
previous models for variable slopes.
They are named $b,c, d$ since they possess the same high density values as the
corresponding $b,c,d$ MDI models~\cite{kubisAPPB41}.  The $e$ model is an
extreme case whose values grow much faster with increasing density than the
rest. Fig.~\ref{EsHighValues} shows $E_s$ for all of them.
\section{Astrophysical observations and Symmetry Energy constraints}

Neutron star models must be compared with astrophysical observations to prove
their validity. Although there are many astrophysical processes,  not all them
allow for precise information about the NS properties, like simultaneous
accurate mass and radius  measurements. However, recently the highest neutron
star mass has been precisely measured~\cite{demorest10} with a value around 2
M$_{\odot}$ placing a strong constraint to many of the existent equations of
state. Even though $E_s$ contributes to the determination of the maximum NS mass
for an EOS it is the isoscalar part $V$ that plays the major role. Therefore
this measurement is not so stringent for $E_s$.  Another pertinent result
regarding NS crusts is the upper bound on the crustal moment of inertia for the
Vela pulsar derived from an analysis performed by Link in~\cite{Link:1999ca},
telling that such quantity should be higher than the 1.4$\%$ of total moment of
inertia of this star. Different $E_s$ forms will result in different crust
properties to be tested against this condition. As a final constraint for the
models presented here the Direct Urca cooling is considered. Low mass NS do not
cool by the DUrca process according to \cite{2006A&A...448..327P}. The proton
fraction of the star $x$ should not go above the DUrca proton fraction threshold
$x_{DU}$ for those low NS masses. The DUrca threshold $x_{DU}$  is weakly
dependent density function and takes values between 0.11 and 0.15.
From the $\beta$-equilibrium equations, Eq.~(\ref{beta}), 
for low values of $x$ and neglecting muons one may get  the following estimation
for the proton fraction in matter
\beq
x = \frac{1}{6+n/n_0 \; (83~\textrm{MeV}/E_s)^3}.
\eeq
This expression implies that one natural way to fulfill this constraint is
restricting $E_s$ to low values, smaller than 80 MeV, so the resulting $x$
always stay below $x_{DU}$. That is the reason we introduce the {\it L-high-b}
models -  the second  sub-class in the {\it L-high} family, which presents soft
behaviour at high density. As we will see in the following sections, this
possibility has important implications for the EOS when satisfying the maximum
NS mass measurement.
 
\section{Neutron star properties}

In order to construct the neutron star profile we need to include an EOS in the
whole range of densities from zero to high densities characteristic of the very
central part of the star. In this study we have considered a liquid core
described by EOS based on the symmetry energy forms presented in the previous
section with a common isoscalar part given by the PAL parametrization. As for
the crust, we have taken the SLy EOS composed of different parts 
whose table can be  found in \cite{ioffe}. These
two EOSs are joint at the point of equal pressure and density. This construction
is not thermodynamically consistent since there is  a discontinuity in chemical
potentials but can be treated as an approximation that does not influence the
overall macroscopic NS properties. In our approach the bottom edge of the crust
is determined by the critical density $n_c(K_{\mu})$ which is very close to
$n_{c} (1\leftrightarrow 2)$. 
The NS total moment of inertia
and that of its crust are calculated by the following expressions, derived in
the framework of General Relativity~\cite{1994ApJ...424..846R}, as:
\begin{eqnarray}
  I& = &\frac{J}{1+2GJ/R^{3}c^{2}}, \nonumber \\
\Delta I_{crust}&=&\frac{2}{3}(M_{crust}R^{2})\frac{1-2GI/R^{3}c^{2}}{1-2GM/Rc^{2}}
\label{IcrustPethick}
\end{eqnarray}
where $M_{crust}=M-M_{core}$ is the difference between the total mass and the mass
of the core, as determined by $n_c$.
Interestingly, the consequences of choosing 
$n_c(Q)$ are systematic effects of lowering the NS crust thickness and its
moment of inertia. The effect was shown in \cite{thesis}.

\subsection{$\Delta E${\it -k} models}

In these models the $E_s$ forms take finite values for very low densities in the
range between $0-12$ MeV sharing the same high density behavior and same $L$
value. They result into an acceptable maximum mass (of about 2
$\textrm{M}_{\odot}$) and radii between $12-13$ kilometers. 
Table~\ref{nccPALDeltaEmodels} shows the
transition densities for  each $k$ value: as $k$ grows the transition density
$n_c$ lowers.
A study of the behavior of the compressibility curves $K_{\mu}$ in this set of
models show that as we increase the values of the symmetry energy at low
densities,  the point of vanishing $K_\mu$ occurs at lower and lower densities
to finally does not appear for k greater than 14~MeV.
This results into thinner and thinner neutron star  crust and
could be extended to crustless neutron stars, which are, however,
not expected to exist according to observations.  
\begin{table}[htbp!]
\caption{Crust-core transition densities ($\textrm{fm}^{-3}$) for the $\Delta
E${\it -k} models.}
\label{nccPALDeltaEmodels}
\begin{tabular}{ccccc}
\hline \hline
  model &$E_s(0)$ &$n_{c}(Q)$ & $n_{c}(K_{\mu})$ & $n_{c} (1\leftrightarrow 2)$ \\
\hline
 $\Delta E$-$k00$ &0& 0.0816675 & 0.0930797 & 0.0942588 \\
 $\Delta E$-$k02$ &2 &0.0690383 & 0.0814262 & 0.0820099 \\
 $\Delta E$-$k04$ &4 &0.053707 & 0.0647734 & 0.0647989 \\
$\Delta E$-$k06$ &6 &0.0399434 & 0.0473238 & 0.0474521\\
$\Delta E$-$k08$ & 8&0.0306162 & 0.0351069 & 0.0353349\\
$\Delta E$-$k10$ &10 &0.0248644 & 0.0278587 & 0.0288262\\
$\Delta E$-$k12$ &12&0.0209877&0.0233639&0.0257871\\
$\Delta E$-$k14$ &14&0.0173067&0.0198872&0.0215231\\
\hline \hline
\end{tabular}
\end{table}
\begin{figure}[htbp!]
\center
\begin{center}
$
\begin{array}{c}
\includegraphics[width=7.5cm]{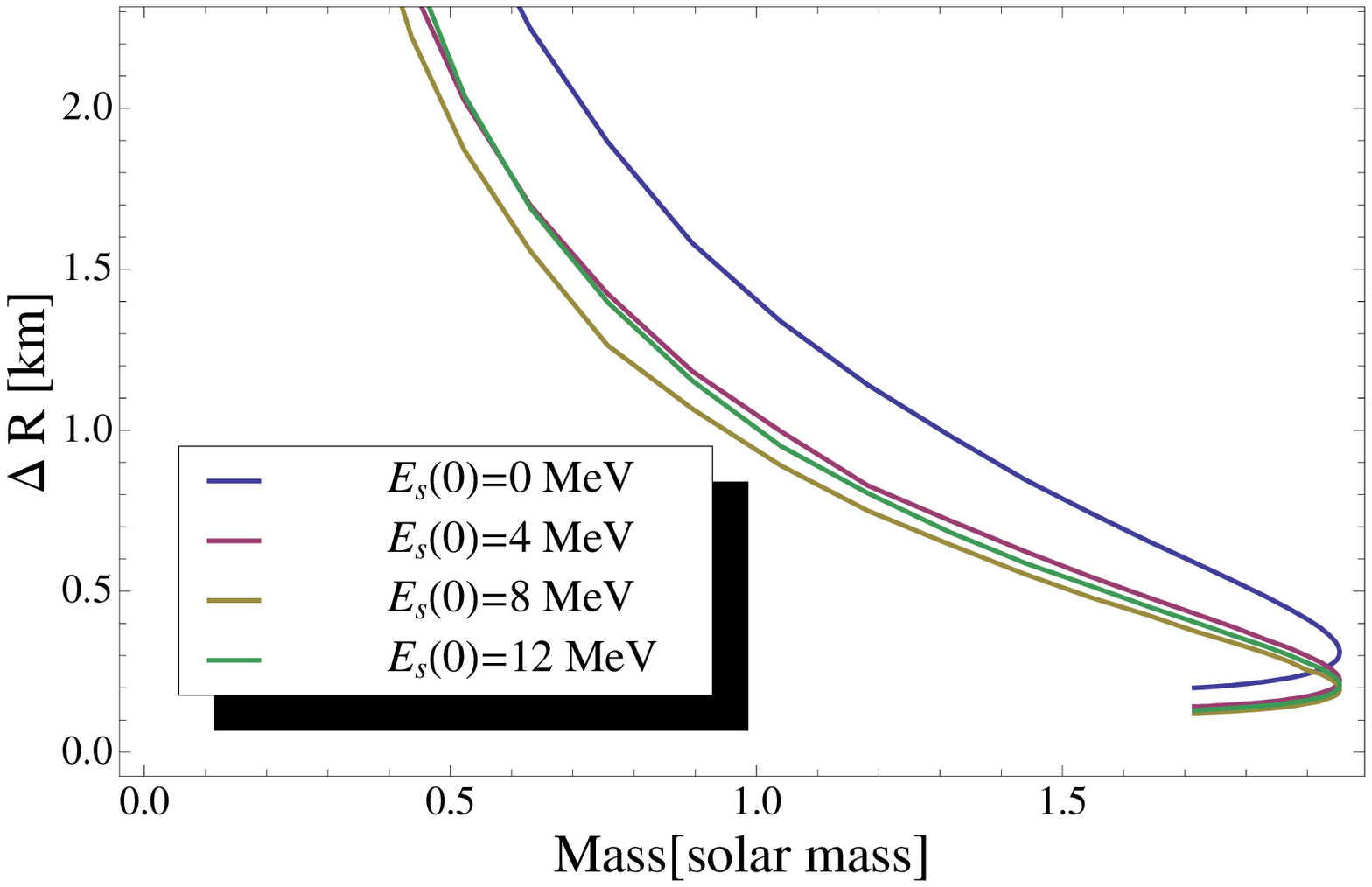}\\
\includegraphics[width=7cm]{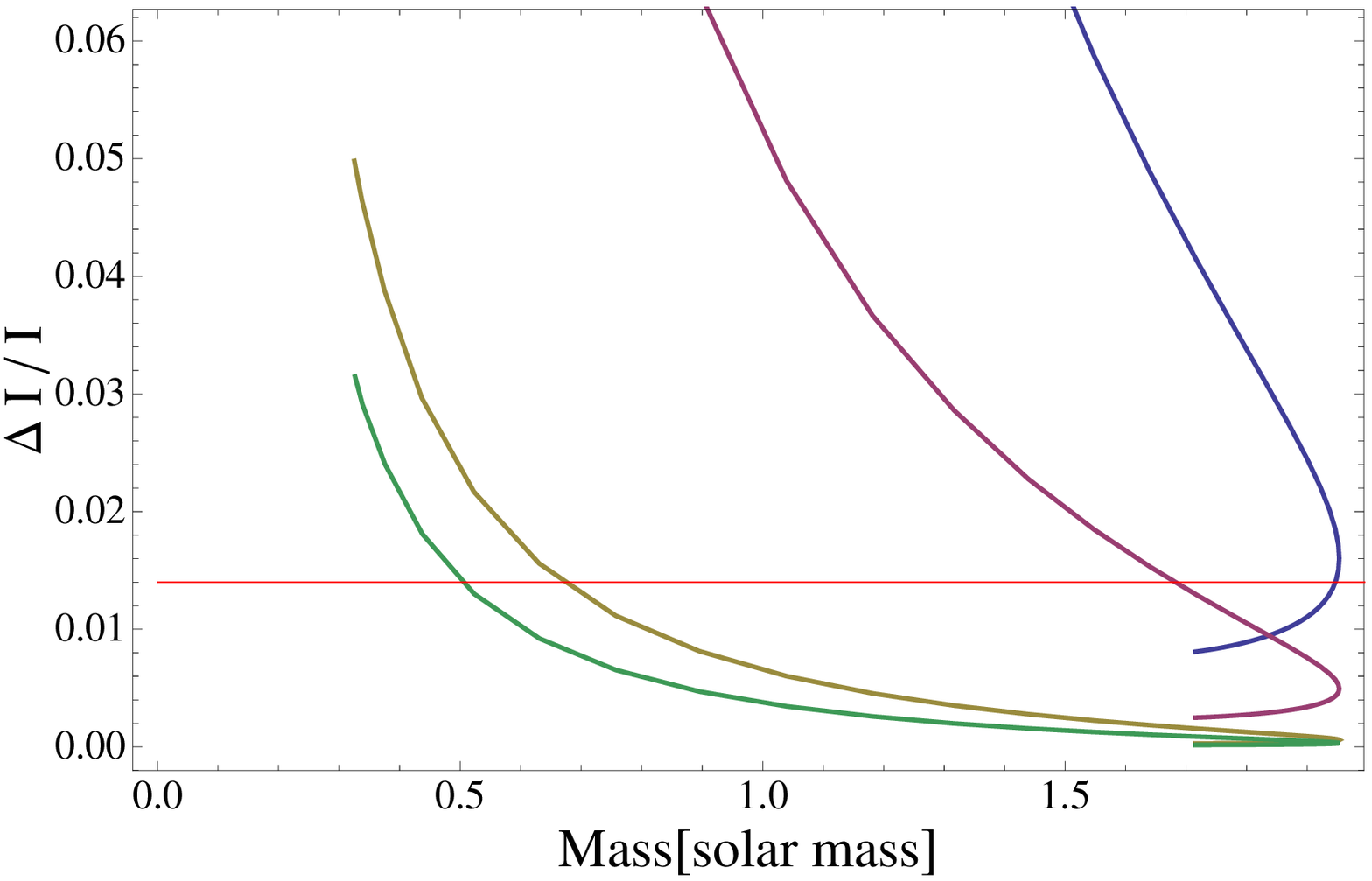}
\end{array}$
\end{center}
\caption{Results for the $\Delta E${\it -k} models. \textit{Top}: crust
thickness. \textit{Bottom}: fractional moment of inertia in the crust where the
red line represent the lower bound for crustal $I$ of the Vela pulsar.
Each curve corresponds to a specific value of $k =E_{s}(0)$.}
\label{PALDeltaE}
\end{figure}
The crustal properties appear to be very sensitive  to the varying portion of
the symmetry energy shape, specially the crustal moment of inertia, which is
shown in Fig.~\ref{PALDeltaE}. The most likely value for the  low density $E_s$
part derived from experiment is around 10 MeV. If it is true, it will point to a
very low mass of the Vela pulsar, much below 1 $\textrm{M}_{\odot}$,  according
to the models here. Such low masses are not favored by the supernova explosion
scenarios which produce new born neutron star with masses around 1.5
$\textrm{M}_{\odot}$. This result is similar to the already reported
in~\cite{kubisAPPB41} but there it could be interpreted as an inconsistency of
the model coming from the clusterization of pure neutron matter. For the models
here this inconsistency has been removed but  the same effect persists.
Summarizing, the above results signal that the large values of $E_s$ at
$n \rightarrow 0$ are in contradiction with neutron star observations.

\subsection{$L$ models}

A first look at the $E_s$ form makes an impression that differences between
different models in the {\it L-high-c} family are not large since they almost
overlap and coincide at low and high density regions. Basically the only
difference between them is the slope at saturation point.
From table \ref{nccLmodels-tab}  we see that the critical density for
crust-core transitions is not clearly correlated to $L$ values: for both high
and low $L$ the critical density is large whereas for the intermediate $L$ is
lower. 
\begin{table}[ht!]
\center
\caption{Crust-core transition densities for the {\it L-high-c} and {\it
L-high-b} models.}
\label{nccLmodels-tab}
\begin{tabular}{cccc}
 \hline \hline
  model & $n_{c}(Q)$ & $n_{c}(K_{\mu})$ & $n_{c}(1\leftrightarrow 2)$ \\
\hline
 L40-high-c & 0.101234 & 0.107808 & 0.113369 \\
 L60-high-c & 0.0918645 & 0.100315 & 0.103118 \\
 L90-high-c & 0.0876315 & 0.100575 & 0.101674 \\
 L120-high-c & 0.111214 & 0.139422 & 0.143345 \\
\hline \hline
 L40-high-b & 0.101234 & 0.107808 & 0.113369 \\
 L60-high-b & 0.0918645 & 0.100315 & 0.103118 \\
 L90-high-b &0.0876315  & 0.100575 & 0.101674 \\
 L120-high-b & 0.111214 & 0.139422 &  0.143345\\
\hline \hline
\end{tabular}
\end{table}
\begin{figure*}[tb!]
\includegraphics[width=7cm]{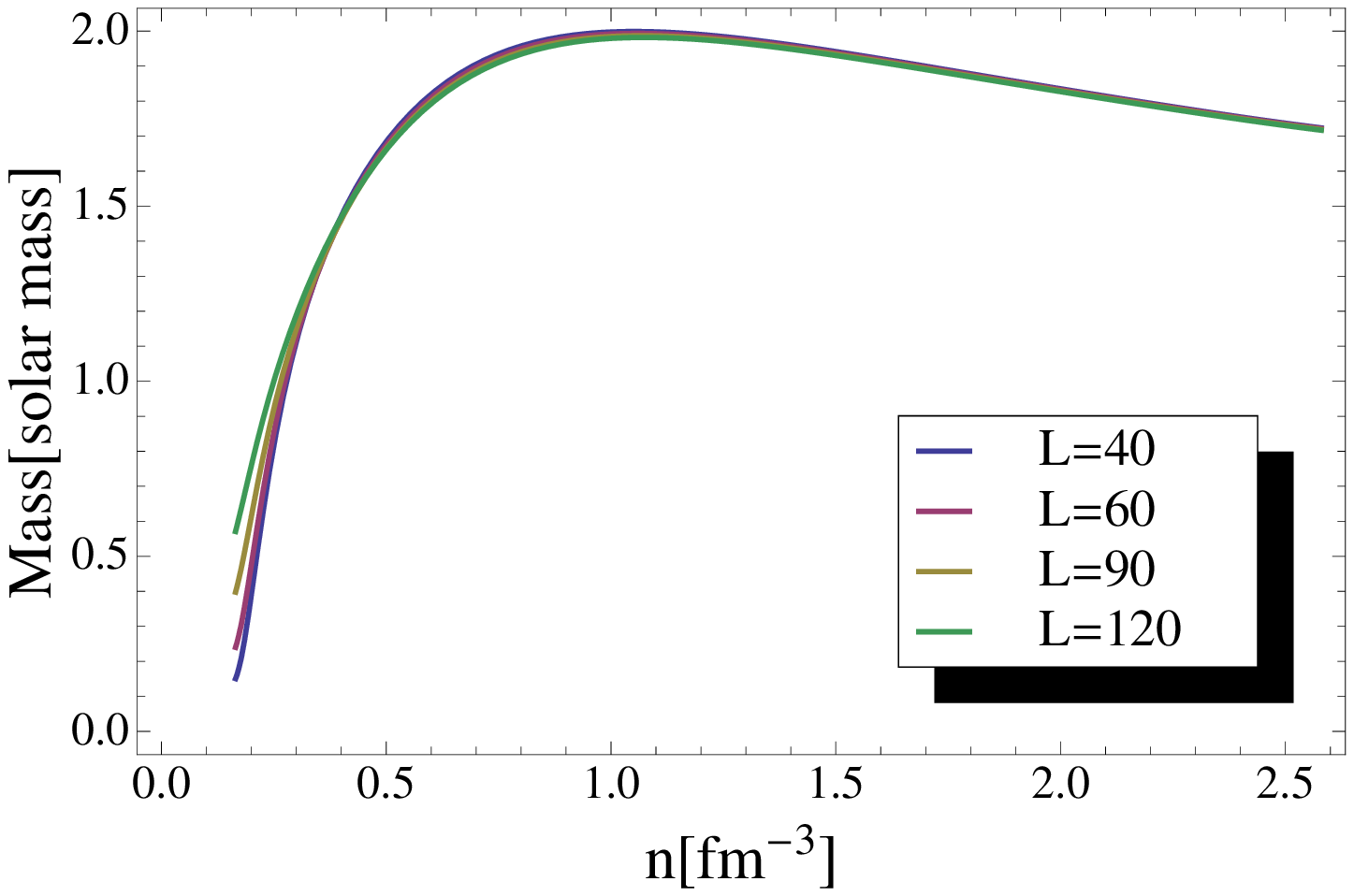}
\includegraphics[width=7cm]{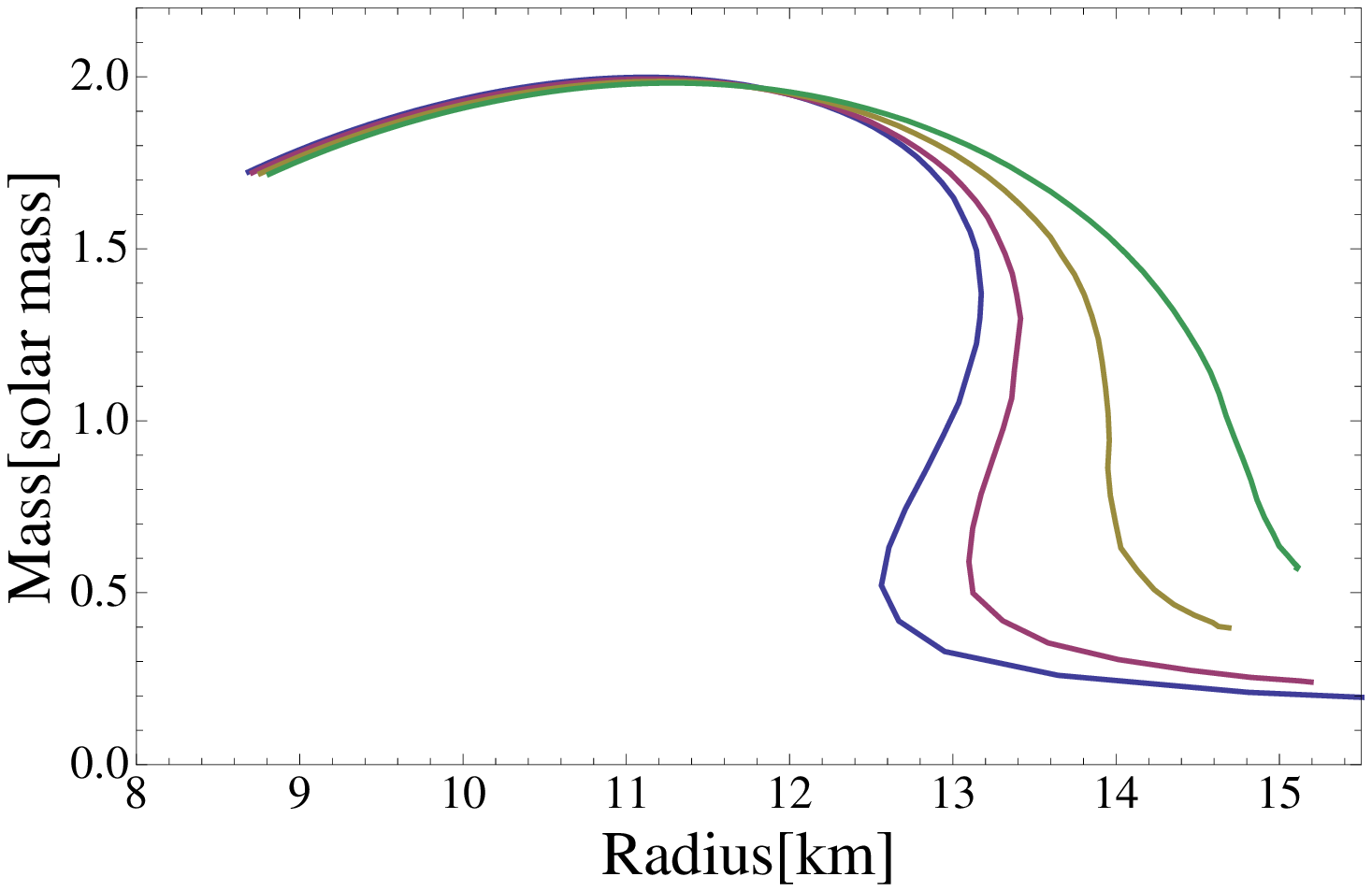}

\includegraphics[width=7cm]{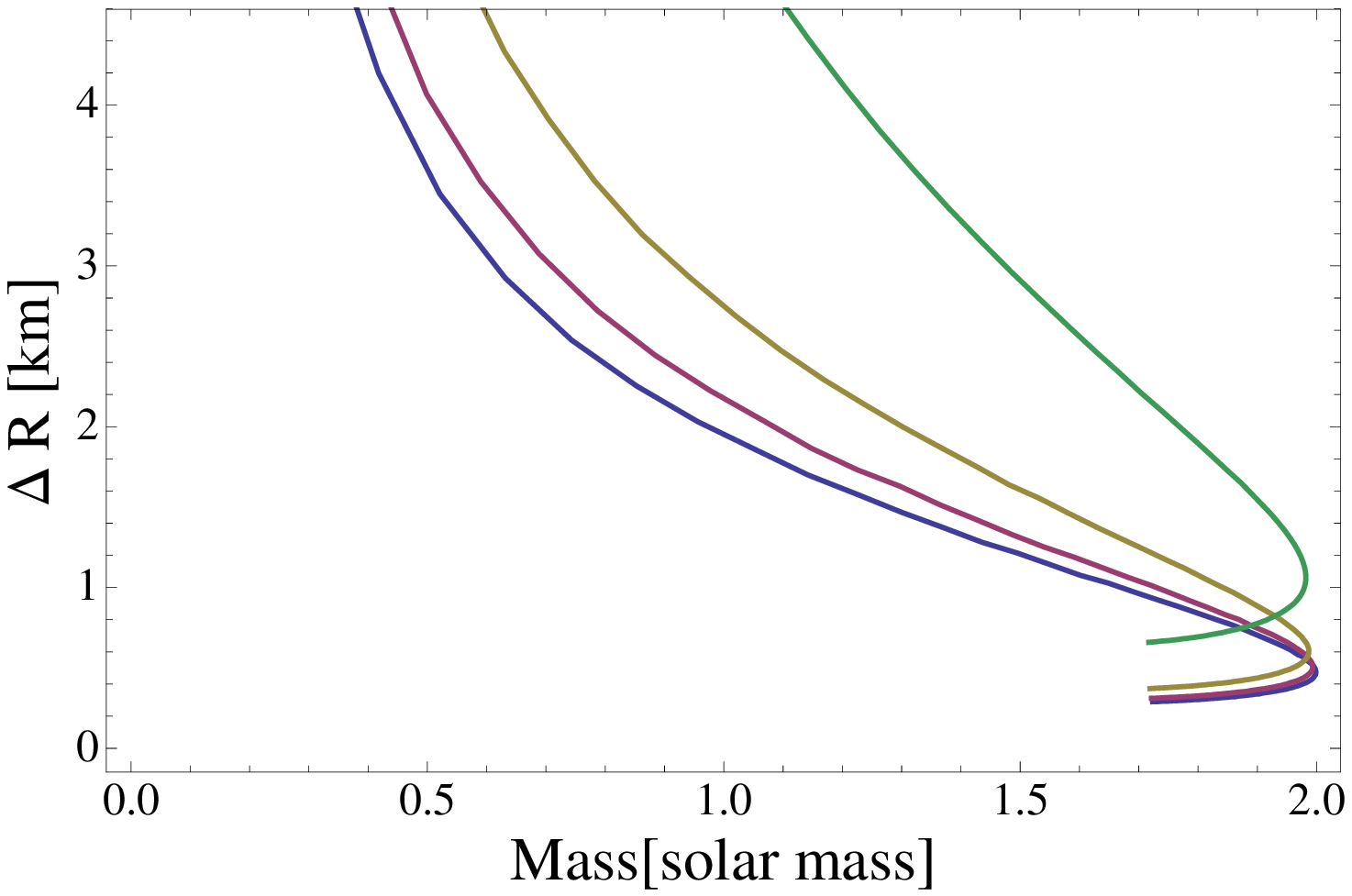}
\includegraphics[width=7cm]{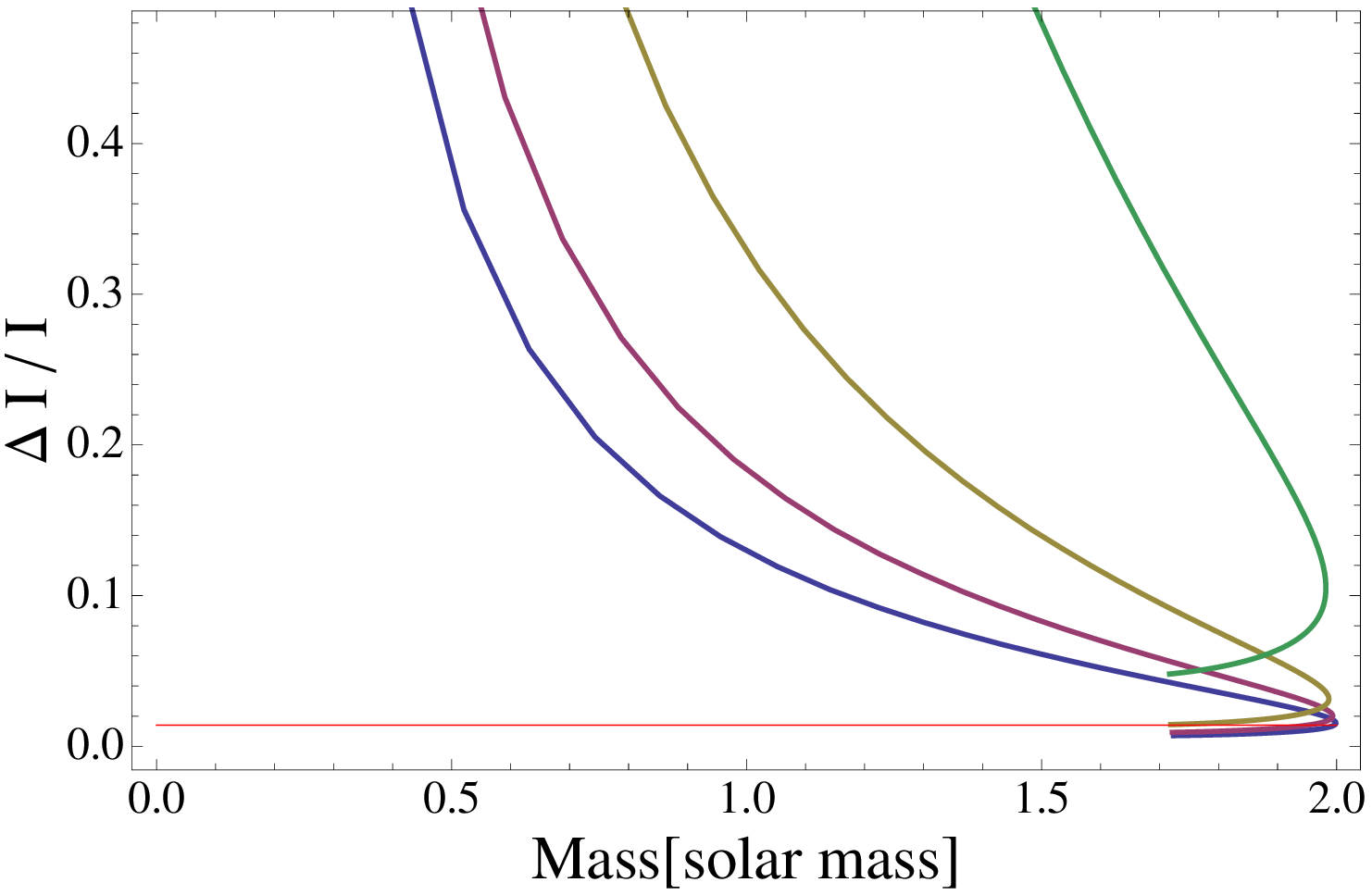}
\caption{Neutron star features for the {\it L-high-c} models.}
\label{Lvar}
\end{figure*}
The behavior of the critical density with $L$ was already analyzed
in~\cite{2009PhRvC..79c5802X} and shows differences when going beyond the
parabolic approximation in the nuclear energy used here. In our approach
critical density differences are not very large but on the contrary the NS
parameters are highly sensitive to the values of $L$.
Both global and crustal properties of NSs are affected by such the different
values of $L$. One may see that the star compactness $\beta\equiv GM/(Rc^{2})$
is essentially changed by $L$ specially for low NS masses. It could be naturally
expected since in low mass NS central densities are relatively low and the low
density $E_s$ values mainly determined by $L$ have to be relevant. The $E_s$
slope contributes to the pressure so increasing
$L$ makes the EOS stiffer and decreases the compactness. On the contrary,  a
soft equation of state amplifies the effect of gravity on the star as a result
of a more compact star. It is interesting that even if the crust-core transition
does not behave monotonically with $L$, the crustal properties like thickness
and moment of inertia depend clearly  on the $E_s$ slope. The larger $L$ is, the
thicker the NS crust becomes. The effect is more pronounced for the moment of
inertia carried by the crust. Small values of $L$ correspond to softer EOS
resulting in a more compact star. A more compact object has stronger gravity
in the crustal region,  therefore the net effect is that the crust is
more squeezed and contributes less to the total moment of inertia. Summarizing,
in the case of $L$ models, the crust-core transition point $n_c$ does not play a
major role in determination of the NS crustal properties but rather the gravity
controlled by $L$ is the main factor here.

\subsection{{\it high-$E_s$} models}

\begin{figure}[ht!]
\center
{\includegraphics[width=7cm]{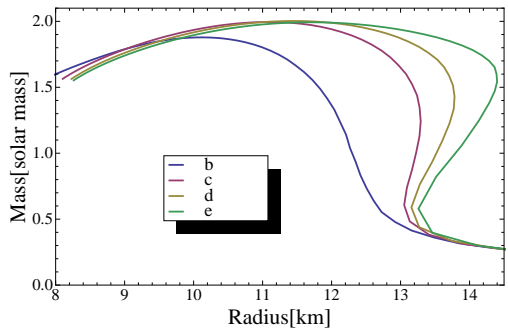}}
\caption{Mass vs radius relation for the {\it high-$E_{s}$} models.}
\label{MvsR-High}
\end{figure}

\begin{figure}[t!]
\begin{center}$
\begin{array}{c}
\includegraphics[width=7cm]{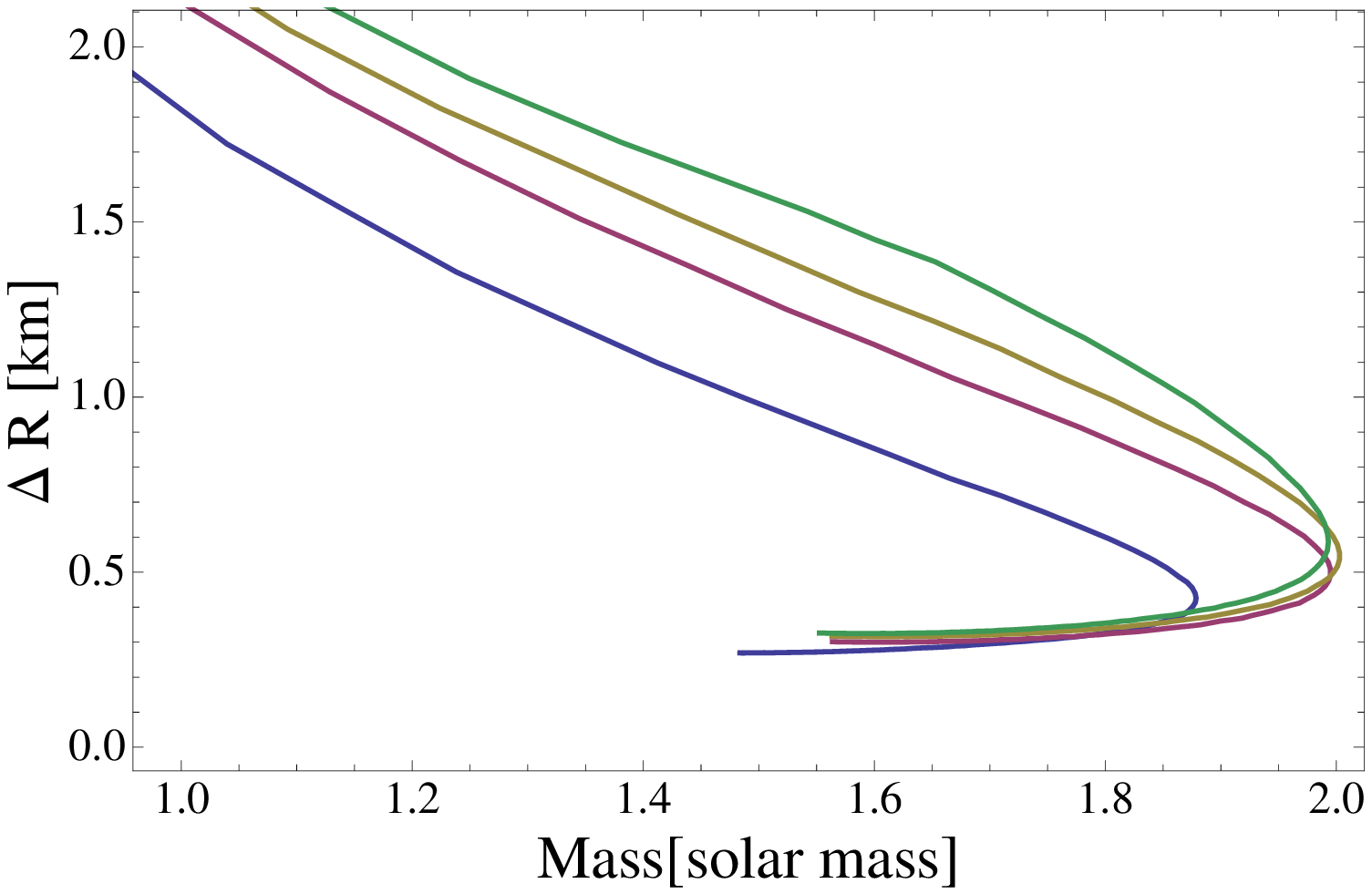}\\
\includegraphics[width=7.3cm]{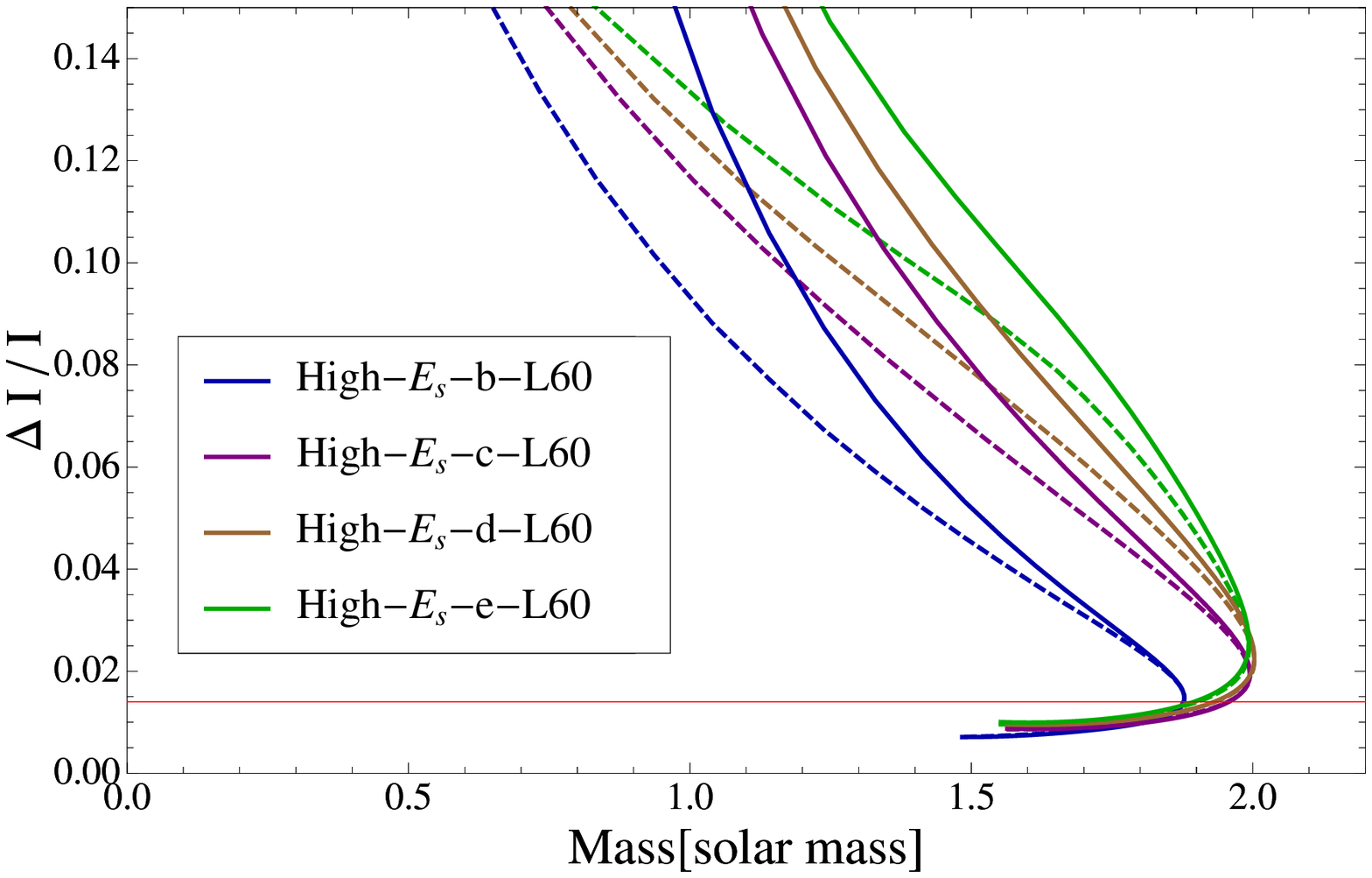}
\end{array}$
\end{center} 
\caption{Effects of the different symmetry energy forms in the {\it
high-$E_{s}$} models for crustal thickness (top) and moment of inertia (bottom).
The dashed  lines correspond to the ${\Delta I_{crust}}/{I}$ derived by the
approximate formula Eq.~(\ref{DeltaIoverI-LP}) }
\label{Highs}
\end{figure}

In this section we present results for the third family. All models in this
family share the same shape of the symmetry energy up to saturation point, which
means that they show the same critical density for the crust-core transition
point, shown in Table~\ref{nccHighEmodels}.
\begin{table}[b!]
\center
\caption{Crust-core transition densities for the {\it high-$E_{s}$}
models.\label{nccHighEmodels}}
\begin{tabular}{cccc}
\hline \hline
 model & $n_{c}(Q)$ & $n_{c}(K_{\mu})$ & $n_{c} (1\leftrightarrow 2)$ \\
\hline
 $high\!-\!E_{s}$ models & 0.0918645 & 0.100315 & 0.103118 \\
\hline \hline
\end{tabular}
\end{table}
Therefore the crustal properties are only affected by the high density shape of
$E_s$.  As one may see from Fig.~\ref{MvsR-High}, the compactness changes a lot
for massive stars in the contrast with the previous family where this property
changed for low massive stars mainly.
The form of the symmetry energy at high densities changes the compactness of the
NS so that the influence of gravity is different in the crust region in
each model. For larger compactness the crust is more squeezed, has lower
thickness and contributes less to the total moment of inertia, what is shown in
Fig.~\ref{Highs}. Those differences are more or less constant and do not depend on
the total mass of a given NS. For moments of inertia the differences are larger
then for the crust thickness. Here we must emphasize that those differences comes
only from the portion of $E_s$ at very high density which is responsible for the
different compactness  of the star. The influence of the compactness on crustal
properties of NS was estimated by an approximate formula for the moment of
inertia derived in~\cite{2001ApJ...550..426L}:
\begin{eqnarray}
\label{DeltaIoverI-LP}
 \frac{\Delta I_{crust}}{I}\simeq\frac{28 \pi P_t R^{3}}{3Mc^{2}}
\frac{(1-1.67\beta-0.6\beta^{2})}{\beta} \nonumber \\
\times \left(1+\frac{2P_t(1+5\beta-14\beta^{2})}{n_t m_b c^{2}\beta^{2}}\right)^{-1}
\end{eqnarray}
where $P_t=p(n_t)$ and $n_t$ (called $n_c$ in this work)  are the  values of
pressure and baryon number density at the crust-core transition and $\beta$ is
the compactness parameter. 
In the Fig.~\ref{Highs} we compare the results of this approximation and we see
discrepancies that show its range of applicability. It means that the 
scaling of the crustal properties with the compactness is not simple and
this formula must be taken with care specially in case of thick crusts.

\subsection{Direct Urca constraint}
\label{sec-durca}

According to \cite{2006A&A...448..327P} neutron stars with masses below 1.35
M$_{\odot}$ should not cool by the direct Urca process.  The proton fraction of
the star $x$ should not go above the DUrca proton fraction threshold $x_{DU}$
for those low NS masses. It is mainly determined by the symmetry energy,
 implying that one way to avoid violating this constraint
is restricting $E_s$ to low values so the resulting $x$ always stay below
$x_{DU}$ (see \cite{Lattimer:1991ib} for detailed discussion).
\begin{figure}
\includegraphics[width=7cm]{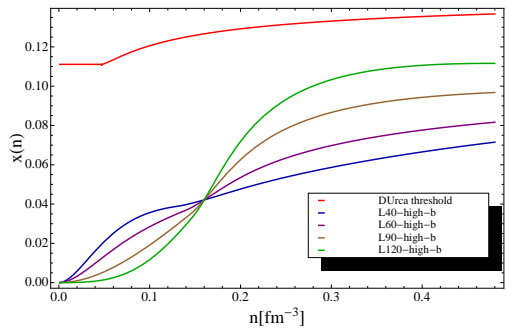}
\caption{The proton fraction and DUrca constraint for the {\it L-high-b} family.}
\label{xDU-L}
\end{figure}
That indeed does not happen for most of the models studied in this work. The
energy expression, Eq.~(\ref{Enuc}), whose symmetric part is based on the PAL
parametrization leads to rather soft EOS, so the central density may be easily
attained by low massive stars. The critical masses satisfying the DUrca
constraint are: 1.2 M$_{\odot}$ for
$\Delta E${\it-k} models, 0.5 - 1.0 M$_{\odot}$ for {\it L-high-c} models  and
0.5 - 1.3 M$_{\odot}$ for \textit{high-$E_s$} models. In the particular case of
the $L$ models the higher density part of $E_s$ is meant to be completely
arbitrary and surely one can incorporate  the cooling limitation easily and
improve them, by means of B\'ezier curves. Figure~\ref{xDU-L} shows the proton
fraction for the {\it L-high-b}
 models whose high density part of
the symmetry energy behaves like the $b$ MDI model. There it can be seen that
the DUrca cooling never sets in since the proton fraction always stays below the
DUrca threshold. This constraint therefore favors $E_s$ forms which do not grow
quickly with increasing density and stay bound from above. However such soft
$E_s$ leads to the maximum mass not greater than 1.85 M$_\odot$ which is in
conflict with observations.

Nevertheless, there exists another possibility apart from modifications to $E_s$
to fulfill the aforementioned restriction. The symmetry energy contributes to
the stiffness of an EOS,  but the main contribution comes from the isoscalar
part of the nuclear energy. As already mentioned, in most of the models with
arbitrary $E_s$ used here  the DUrca cooling sets in for unacceptable masses,
due to the particular form of $E_s$.  But if it turns out that $E_s$ is such
that $x$ never goes above the DUrca threshold $x_{DU}$ (like in the {\it
L-high-b} models) then it's most likely that the isoscalar part of the PAL
parametrization be wrong. It should be then properly replaced by an expression
that enhances the stiffness of the EOS that would not only satisfy the DUrca
constraint but at the same time contribute to the appearance of high mass
neutron stars. It is worth mentioning that even if the PAL formula is highly
biased, it has been useful to spot the $E_s$ contribution to the NS physics.  On
the contrary, if a PAL-like isoscalar part is correct,  then the symmetry
energy  can be very much constrained by both the cooling and the maximum mass,
resulting in moderate values at low densities (to satisfy DUrca)  and then
increasing abruptly  so high mass NS are created within the model. This result
could be in fact very stringent,  and its validity is to be studied and
confirmed with the upcoming observations. As a final remark it is important to
mention that future approaches to the EOS should take the cooling phenomenon as
an important test. 

\section{Conclusions}

In this work we were interested in the role of the symmetry energy form  played
in the NS crust properties. The crust-core transition occurs well below the
saturation density, so the present knowledge of the $E_s$ parameters at this
density is not sufficient to determine the transition point  exactly. Moreover
the symmetry energy form at higher densities, although not affecting the 
crust-core transition point directly is also relevant. One may say the whole
shape of the $E_s$ has to be taken into account. In order to extract the
influence of different portions of the $E_s$  shape we have constructed the
three different families of symmetry energy parametrization corresponding the
three range of densities: well below the saturation point, around $n_0$ and
highly above. In each family one chosen feature was changing whereas the
remaining portion of $E_s(n)$ was kept the same. This was achieved
by use of B\'ezier curves parametrization.

In the low density regime ($\Delta E$-{\it k}-models) the characteristic feature was the
value of $E_s(n\rightarrow 0)$ being  not 0 as it is suggested by some recent
measurement. It appeared that high values of $E_s$ at low density are
questionable as they lead to a very thin crust, difficult to be reconciled with
observations of the Vela pulsar. 

In the intermediate range of density {\it L}-models the key quantity was the slope of symmetry
energy $L$ taken in the range from 40 to 120 MeV. The crust thickness and 
moment of inertia  is very sensitive and highly increases with the value of $L$.

For the high densities we constructed the models preserving the same crust-core
transition density but presenting completely different behaviour well above 
$n_0$, they were called {\it high-$E_s$} models. It appeared that values of
$E_s$ at high density affect the crustal properties essentially. Such
effect comes from different compactness of a star. However it is difficult to
find a simple scaling of crust properties with $M/R$, especially when the crust
is thick.

The direct Urca constraint for low mass NS is related to the symmetry energy since it serves to determine the proton fraction inside the star.
If the high density behavior of $E_s$ is bounded from above to satisfy this constraint a 2 M$_\odot$ NS cannot be created by the models used here.
A stiffer isoscalar part $V$ of the nuclear energy per particle is then necessary to produce higher masses. Therefore both the heaviest observed NS and the
DUrca cooling condition allow for constraining the EOS, since they are to be satisfied simultaneously. Solving this issue is surely an interesting implementation 
for a future work.

During the preparation of the text the authors have found the work by Lattimer
et al.~\cite{2012arXiv1203.4286L} which  combines the majority of laboratory
measurements. It is concluded there that the most reliable values of $S_v
\approx 32 ~\textrm{MeV},~~L\approx 50 ~\textrm{MeV}$ with an error of a
few MeV. This means that $L$-dependence on the crustal properties is almost
removed. In our work, to avoid additional complexity,  we have fixed the second
derivative $K_s = 0$ for most presented models. Now, when the first derivative
is pinned down  it seems interesting to explore different $K_s$ values, what is
planned for the future.

\begin{acknowledgements}
This work has been partially supported by CompStar a research networking
programme of the European Science Foundation.
The authors are grateful to D. Blaschke for illuminating discussions and talks
on the symmetry energy subject. 
\end{acknowledgements}

\appendix*

\section{B\'ezier points for various models}

\begin{table}[ht!]
\label{DeltaE-k-modelPoints}
\caption{B\'ezier control points for the $\Delta E-k$ models. $\Delta E -k00$ is simply the PALu model.}
\begin{tabular}{cccccc}
 \hline \hline
  model & $\textbf{P}_0$ & $\textbf{P}_1$ & $\textbf{P}_2$ & $\textbf{P}_3$ & $\textbf{P}_4$\\
\hline
 $\Delta E -k02$ & (0, 2) & (0.04, 1) & (0.08, 0) & (0.12, 0) & (0.16, 0)  \\
 $\Delta E -k04$ & (0, 4)& (0.04, 2) & (0.08, 0) & (0.12, 0) & (0.16, 0) \\
 $\Delta E -k06$ & (0, 6) & (0.04, 3) & (0.08, 0) & (0.12, 0) & (0.16, 0) \\
 $\Delta E -k08$ & (0, 8) & (0.04, 4) & (0.08, 0) & (0.12, 0) & (0.16, 0) \\
 $\Delta E -k10$ & (0, 10) & (0.04, 5) & (0.08, 0) & (0.12, 0) & (0.16, 0) \\
 $\Delta E -k12$ & (0, 12) & (0.04, 6) & (0.08, 0) & (0.12, 0) & (0.16, 0) \\
 $\Delta E -k14$ & (0, 14) & (0.04, 7) & (0.08, 0) & (0.12, 0) & (0.16, 0) \\
\hline \hline
\end{tabular}
\end{table}

\begin{table*}[htpb!]
\label{LmodelPoints}
\caption{B\'ezier control points for the $L$ models which are composed of two B\'ezier curves joint at $n_0$.}
{\begin{tabular}{ccccccc}
 \hline \hline
  model & $\textbf{P}_0$ & $\textbf{P}_1$ & $\textbf{P}_2$ & $\textbf{P}_3$ & $\textbf{P}_4$ & $\textbf{P}_5$\\
\hline
L40-high-c& (0, 0) & (0.0528, 22.066) & (0.106, 26.466) & (0.16, 31) &  &  \\
(low density)&&&&&&\\
 L40-high-c & (0.16, 31) & (0.24, 37.666) & (0.32, 44.333) & (0.8, 235.372) & (1.28, 455.583) & (1.92, 819.168) \\
(high density)&&&&&&\\ 
\hline
L60-high-c & (0, 0) & (0.0528, 17.6) & (0.106, 24.2) & (0.16, 31) &  & \\
 & (0.16, 31) & (0.24, 41) & (0.32, 51) & (0.8, 235.372) & (1.28, 455.584) & (1.92, 819.168) \\
\hline
L90-high-c & (0, 0) & (0.0523, 10.9) & (0.106, 20.8) & (0.16, 31) &  & \\
 &(0.16, 31) & (0.24, 469) & (0.32, 61) & (0.8, 235.372) & (1.28, 455.584) & (1.92, 819.168) \\
\hline
L120-high-c & (0,0) & (0.053, 4.2) & (0.106, 17.4) & (0.16, 31) &  &  \\
 & (0.16, 31) & (0.24, 51) & (0.32, 71) & (0.8, 235.372) & (1.28, 455.584) & (1.92, 819.168) \\
\hline
\hline
L40-high-b& (0, 0) & (0.0528, 22.066) & (0.106, 26.466) & (0.16, 31) &  &  \\
 & (0.16, 31) & (0.24, 37.666) & (0.32, 44.333) & (0.8, 72.239) & (1.28, 79.064) & (1.92, 71.955) \\
\hline
L60-high-b & (0, 0) & (0.0528, 17.6) & (0.106, 24.2) & (0.16, 31) &  & \\
 & (0.16, 31) & (0.24, 41.000) & (0.32, 51.000) & (0.8, 72.239) & (1.28, 79.064) & (1.92, 71.955) \\
\hline
L90-high-b & (0, 0) & (0.0523, 10.9) & (0.106, 20.8) & (0.16, 31) &  & \\
 &(0.16, 31) & (0.24, 46.000) & (0.32, 61) & (0.8,72.239 ) & (1.28, 79.064) & (1.92, 71.955) \\
\hline
L120-high-b & (0,0) & (0.053, 4.2) & (0.106, 17.4) & (0.16, 31) &  &  \\
 & (0.16, 31) & (0.24, 51) & (0.32, 71) & (0.8, 72.239) & (1.28, 79.064) & (1.92, 71.955) \\
\hline \hline
\end{tabular}}
\end{table*}

\begin{table*}[hpbt!]
\label{hmodelPoints}
\caption{B\'ezier control points for the $high-E_s$ models.}
\center
{\begin{tabular}{ccccccc}
 \hline \hline
  model & $\textbf{P}_0$ & $\textbf{P}_1$ & $\textbf{P}_2$ & $\textbf{P}_3$ & $\textbf{P}_4$ & $\textbf{P}_5$\\
\hline
low density & (0, 0) & (0.0528, 17.6) & (0.106, 24.2) & (0.16, 31) &  & \\
\hline
high density   &&&&&&\\ 
\hline
L60-high-b& (0.16, 31) & (0.24, 41) & (0.32, 51) & (0.8, 89.12) & (1.28, 82.86) & (1.6, 71.92) \\
L60-high-c & (0.16, 31) & (0.24, 41) & (0.32, 51) & (0.8, 207.26) & (1.28, 440.014) & (1.6, 626.88) \\
L60-high-d & (0.16, 31) & (0.24, 41) & (0.32, 51) & (0.8, 301.097) & (1.28, 813.098) & (1.6, 1146.13) \\
L60-high-e & (0.16, 31) & (0.24, 41) & (0.32, 51) & (0.8, 547.93) & (1.28, 1244.98) & (1.6, 1673.62) \\
\hline \hline
\end{tabular}}
\end{table*}

\newcommand{\physrep}{Phys.~Rep.~}
\newcommand{\aap}{A\&A}


\begin{thebibliography}{99} 
\bibitem{Kubis:2006kb}
  S.~Kubis,
  Phys.\ Rev.\  C {\bf 76} (2007) 025801
\bibitem[Kubis et al.(2009)]{kubisAPPB41} Kubis, S., Porebska, J.,\&
Alvarez-Castillo, D.~E.\ Acta Phys.Polon.B41:2449,2010
\bibitem[Natowitz et al.(2010)]{2010PhRvL.104t2501N} Natowitz, J.~B., et al.\
2010, Physical Review Letters, 104, 202501 
\bibitem[Baym et al.(1971)]{1971NuPhA.175..225B} Baym, G., Bethe, H.~A., \&
Pethick, C.~J.\ 1971, Nuclear Physics A, 175, 225 
\bibitem[Myers \& Swiatecki(1974)]{1974AnPhy..84..186M} Myers, W.~D., \&
Swiatecki, W.~J.\ 1974, Annals of Physics, 84, 186
\bibitem{Chen:2010qx} 
  L.~-W.~Chen, C.~M.~Ko, B.~-A.~Li and J.~Xu,
  Phys.\ Rev.\ C {\bf 82}, 024321 (2010)

\bibitem{Tsang:2008fd} 
  M.~B.~Tsang, Y.~Zhang, P.~Danielewicz, M.~Famiano, Z.~Li, W.~G.~Lynch and A.~W.~Steiner,
  Phys.\ Rev.\ Lett.\  {\bf 102}, 122701 (2009)

\bibitem{Chen:2004si} 
  L.~-W.~Chen, C.~M.~Ko and B.~-A.~Li,
  Phys.\ Rev.\ Lett.\  {\bf 94}, 032701 (2005)

\bibitem{Centelles:2008vu} 
  M.~Centelles, X.~Roca-Maza, X.~Vinas and M.~Warda,
  Phys.\ Rev.\ Lett.\  {\bf 102}, 122502 (2009)
  [arXiv:0806.2886 [nucl-th]].

\bibitem{Li:2007bp} 
  T.~Li, U.~Garg, Y.~Liu, R.~Marks, B.~K.~Nayak, P.~V.~M.~Rao, M.~Fujiwara and H.~Hashimoto {\it et al.},
  Phys.\ Rev.\ Lett.\  {\bf 99}, 162503 (2007)
  [arXiv:0709.0567 [nucl-ex]].

\bibitem{Kowalski:2006ju} 
  S.~Kowalski, J.~B.~Natowitz, S.~Shlomo, R.~Wada, K.~Hagel, J.~Wang, T.~Materna and Z.~Chen {\it et al.},
  Phys.\ Rev.\ C {\bf 75}, 014601 (2007)
  [nucl-ex/0602023].


\bibitem[]{BezierCurves} http://en.wikipedia.org/wiki/Bezier\_curve \\
Farin, Gerald, Curves and surfaces for computer-aided geometric design (4 ed.), Elsevier Science \& Technology Books, 1997
\bibitem[Prakash et al.(1988)]{1988PhRvL..61.2518P} Prakash, M., Lattimer,
J.~M., \& Ainsworth, T.~L.\ 1988, Physical Review Letters, 61, 2518
\bibitem[Lattimer \& Prakash(2001)]{2001ApJ...550..426L} Lattimer, J.~M., \& Prakash, M.\ 2001, \apj, 550, 426
\bibitem{demorest10} P. B. Demorest, T. Pennucci, S. M. Ransom, M. S. E. Roberts and J. W. T. Hessels, {\it Nature}, {\bf 467}, 1081 (2010).
\bibitem{ioffe} http://www.ioffe.ru/astro/NSG/NSEOS/
\bibitem{thesis} D.E. Alvarez-Castillo, PhD Thesis, 2012.X
\bibitem{Link:1999ca}
  B.~Link, R.~I.~Epstein and J.~M.~Lattimer,
  Phys.\ Rev.\ Lett.\  {\bf 83} (1999) 3362
\bibitem[Ravenhall \& Pethick(1994)]{1994ApJ...424..846R} Ravenhall, D.~G., \& Pethick, C.~J.\ 1994, \apj, 424, 846
\bibitem[Popov et al.(2006)]{2006A&A...448..327P} Popov, S., Grigorian, H., Turolla, R., \& Blaschke, D.\ 2006, \aap, 448, 327 
\bibitem{Chen:2007ih} 
  L.~-W.~Chen, C.~M.~Ko and B.~-A.~Li,
  Phys.\ Rev.\ C {\bf 76}, 054316 (2007)
\bibitem[Xu et al.(2009)]{2009PhRvC..79c5802X} Xu, J., Chen, L.-W., Li, B.-A., \& Ma, H.-R.\ 2009, \prc, 79, 035802
\bibitem{Lattimer:1991ib} 
  J.~M.~Lattimer, M.~Prakash, C.~J.~Pethick and P.~Haensel,
  Phys.\ Rev.\ Lett.\  {\bf 66}, 2701 (1991).

\bibitem[Lattimer \& Lim(2012)]{2012arXiv1203.4286L} Lattimer, J.~M., \& Lim,
Y.\ 2012, arXiv:1203.4286 
\end{thebibliography}
\end{document}